\documentclass[reprint, prb, notitlepage]{revtex4-1}

\usepackage{color}
\usepackage{graphicx}
\usepackage{amssymb, amsmath}
\begin{document}

\title{
Cumulant methods for electron-phonon problems. I. Perturbative expansions}

\author{Paul J. Robinson}
\thanks{These two authors contributed equally}
\author{Ian S. Dunn}
\thanks{These two authors contributed equally}
\author{David R. Reichman}
\email{drr2103@columbia.edu }
\affiliation{Department of Chemistry, Columbia University, New York, New York 10027, United States}

\begin{abstract}
In this work we investigate the ability of the cumulant expansion (CE) to capture one-particle spectral information in electron-phonon coupled systems at both zero and finite temperatures. In particular, we present a comprehensive study of the second- and fourth-order CE for the one-dimensional Holstein model as compared with numerically exact methods.  We investigate both finite sized systems as well as the approach to the thermodynamic limit, drawing distinctions and connections between the behavior of systems in and away from the thermodynamic limit that enable a greater understanding of the ability of the CE to capture real-frequency information across the full range of wave vectors.
We find that for zero electronic momentum, the spectral function is well described by the second-order CE at low and high temperatures. However, for non-zero electronic momenta, the CE is only accurate at high temperature. We analyze the fourth-order cumulant, and find that while it improves the description of the short-time dynamics encoded in the one-particle Green's function, it can introduce divergences in the time domain as well as unphysical negative spectral weight in the spectral function. When well-behaved, the fourth-order CE does provide notable accurate corrections to the second-order CE.  Finally, we use our results to comment on the use of the CE as a tool for calculating transport behavior in the realistic {\em ab initio} modeling of materials.
\end{abstract}

\maketitle
The description of the dynamics of electrons interacting with phonons is a cornerstone
topic in condensed matter physics due to its ubiquity and the importance of electron-phonon 
interactions (EPIs) in determining the properties of solids.
Indeed, EPIs are crucial for understanding a wide range of
phenomena in solids, including superconductivity, transport properties, and the vibronic 
satellite structure in emission and absorption spectra, to name 
just a few.\cite{Bardeen1957, Wang2011, Mahan2000, Mishchenko2015, Song2015, Spano2010,Oleson2019}
Unfortunately, even for simplified canonical EPI models, such as the 
Holstein, Fr{\" o}hlich, and Su-Schrieffer-Heeger models
which were introduced many decades ago, 
exact dynamical solutions are
largely out of reach.
\cite{Holstein1959a, Holstein1959, Frohlich1954, Su1979}

There is a plethora
of methods for extracting 
accurate properties of EPI models which may be useful under different circumstances.
Exact ground state and low-lying excited state properties
for the Holstein and other models are attainable through diagonalization in a
variational Hilbert space (VD). \cite{Trugman1990,Bonca1999,Ku2002}
Focusing on the electron-phonon dynamics, the one-particle Green's function
$\mathcal{G}(k,t)$ has been extensively studied at zero-temperature using 
exact diagonalization,\cite{Ranninger1992, Marsiglio1993, DeMello1997, Fehske1997}
cluster perturbation theory,\cite{Alexandrov1994,Hohenadler2003}
a variational approach,\cite{DeFilippis2005}
the momentum-averaged approximation,\cite{Berciu2006,Goodvin2006}
and diagrammatic quantum Monte
Carlo (DQMC) in conjunction with
numerical analytic continuation.\cite{Mishchenko2000}

The frontier of finite-temperature
dynamical calculations remains less well explored. The increased
occupation of higher-lying phonon states at nonzero temperatures renders
Fock space methods harder to converge. 
Only recently has the spectral function ($A\left(k,\omega\right) = -\pi^{-1}\Im \mathcal{G}(k,\omega)$)
been reported at finite temperature for the Holstein model 
using VD with the finite-temperature Lanczos method
on 6- and 12-site systems. \cite{Jaklic2000, Prelovsek2013,Bonca2019} 
DQMC has provided the temperature
dependent mobilities for the Holstein and Fr{\"o}hlich models;
however, as with spectral information this approach is restricted by an ill-conditioned analytic continuation
procedure.\cite{Mishchenko2015, Mishchenko2018}
More recently, numerically exact dynamical methods based on DMRG+VD,\cite{Jansen2020} 
a generalized cluster expansion,\cite{Carbone2021} and the Hierarchical Equations of Motion (HEOM)\cite{Tanimura1989,Ishizaki2005,Tanimura2006,Ishizaki2009,Shi2009,Liu2014,Chen2015,Dunn2019,jankovic2021spectral} approach
have been introduced for real-time dynamics in 
lattice models with EPIs.

For realistic \textit{ab initio} modeling\cite{Luders2005,Giustino2017,Verdi2017,Nery2018,Sio2019,Sio2019a,Zhou2019Predicting,chang2021intermediate}
of systems with EPIs, many of the exact methods mentioned previously are infeasible. Instead, perturbative approaches are usually employed.
However, since
each order of perturbation theory exponentially increases the number of self-energy diagrams, 
it is not practical or computationally efficient to directly compute high-order diagrams in large, realistic systems, 
and approximate resummations of higher-order terms become essential.
The cumulant expansion (CE) approach has been used for this purpose for many years,
and was recently combined with 
density functional perturbation theory to calculate the 
finite temperature photoemission spectra of MgO,
LiF \cite{Nery2018}, and TiO\textsubscript{2}. \cite{Verdi2017}
While the utility of the CE for the calculation of $\mathcal{G}\left(k,t\right)$ at finite temperature has been 
known for years, only a few papers have systematically explored its validity.\cite{Dunn1975, mahan1966phonon,Gunnarsson1994} 

Motivated by recent exact dynamical results in the finite-temperature Holstein model,\cite{Bonca2019}
we systematically explore the CE in this system as proposed by Dunn in the context of Fr{\" o}hlich insulators. \cite{Dunn1975} 
A similar expansion was also used in conjunction
with the Matsubara formalism by Gunnarsson \textit{et al} for describing
{\em zero-temperature} spectral properties of the half-filled Holstein model.  
\cite{Gunnarsson1994}
In Section \ref{sec:Theory} we introduce the model, as well as
the definition of $\mathcal{G}\left(k,t\right)$, and the framework of the CE.
In Section \ref{sec:Results} we provide a detailed comparison of the CE in the 6-site Holstein model
with exact VD results. This comparison highlights a number of interesting features which demand 
more detailed investigation.
In Section \ref{sec:cumulant_finite_size} we discuss errors of the CE that are associated with finite lattice size.
In Section
\ref{subsubsec:Divergences-in-the} we demonstrate several useful and problematic features of the fourth-order CE.
In Section \ref{sec:time_convergence} we analyze the short-time convergence of
the CE as well as the limitations of the CE in capturing fine spectral features and long-time behavior. 
In Section \ref{sec:thermodynamic_limit} we will present results
for the spectral function of an infinite system.  We conclude with a summary of the main results and the implications of our findings for the use of the CE in the {\em ab initio} modeling of materials.

\section{Model and Perturbative Cumulant Expansion\label{sec:Theory}}
In this section we provide background information needed for the remainder of the paper.
While all of the information in this section is well-known, this information is useful
for setting notation and for providing a self-contained discussion of the results that follow.
Throughout this work we focus only on a very specific model, namely the one-dimensional Holstein model with Einstein phonons and periodic boundary conditions. \cite{Holstein1959a,Holstein1959,Mahan2000}
We consider only the single particle case, that is a single electron promoted into an otherwise empty band.
The model is defined by a system-bath Hamiltonian
\begin{align}
H & =H_{e}+H_{p}+V,\label{eq:1}
\end{align}
where the kinetic energy term
\begin{align}
H_{e} & \equiv-t_0\sum_{n}\left(a_{n}^{\dagger}a_{n+1}+a_{n}^{\dagger}a_{n-1}\right)=\sum_{k}\varepsilon_{k}a_{k}^{\dagger}a_{k},\label{eq:-3.6}\\
\varepsilon_{k} & =-2t_0\cos k,
\end{align}
describes the purely electronic system, and
\begin{align}
H_{p} & \equiv\omega_{0}\sum_{n}b_{n}^{\dagger}b_{n}=\omega_{0}\sum_{k}b_{k}^{\dagger}b_{k},
\end{align}
describes the bath. Lastly,
\begin{align}
V & \equiv g\omega_{0}\sum_{n}a_{n}^{\dagger}a_{n}\left(b_{n}+b_{n}^{\dagger}\right)
\nonumber \\
& =\frac{g\omega_{0}}{\sqrt{N}}\sum_{kq}a_{k+q}^{\dagger}a_{k}\left(b_{q}+b_{-q}^{\dagger}\right),\label{eq:-3.7}
\end{align}
accounts for the EPI, which is linear in the bath coordinates. The Holstein
model 
describes the deformation of a discrete lattice, \cite{Holstein1959a,Holstein1959} reflecting
the decoupled nature of sites in a molecular crystal by including
only strictly local electron-phonon coupling. In addition, the model further isolates the effects of intermolecular relaxation by ignoring Peierls-like coupling. \cite{Bariic1970,Su1979,Sous2018,Sous2018a}
For an excellent review that discusses the relation between the Holstein
model and continuum models such as the Fr{\" o}hlich model, see the work of Devreese and Alexandrov.
\cite{Devreese2009}

\subsection{One-Particle Green's Function\label{subsec:One-Particle-Green's-Function}}
We will focus on the calculation of the finite temperature one-particle 
(causal
\footnote{In the model studied here with a single electron, this is equivalent to the retarded Green's function})
Green's function, \cite{Mahan2000,Fetter2003}
\begin{align}
\mathcal{G}\left(k,t\right) & \equiv-i\Theta\left(t\right)\frac{\text{Tr}\left[e^{-\beta\left(H-\mu N\right)}a_{k}(t)a_{k}^{\dagger}(0)\right]}{\text{Tr}\left[e^{-\beta\left(H-\mu N\right)}\right]}.\label{eq:-19}
\end{align}
This quantity is directly related to experimentally measurable quasi-particle spectra as probed by, e.g., photoemission spectroscopy, and can be used to infer transport properties such as charge mobilities in an approximate manner.
\cite{Mahan2000, Mahan1966, chang2021intermediate, Zhou2019Predicting}
In addition, the one-particle Green's function provides a testbed for the comparison of numerical methods ranging from the approximate to the exact which may be applied to general electron-phonon problems. \cite{Damascelli2003,Inosov2008,Mishchenko2000,Mishchenko2018}

As mentioned above, we study an insulator
where the chemical potential $\mu$ satisfies $\mu\ll-2|t_0|$ and there is a single electron placed in the conduction band.\cite{Dunn1975}
When this is the case, it is simple to demonstrate that the trace over the many-electron Fock space in Eq. \ref{eq:-19} can be exactly replaced by a trace over zero-electron
states and non-interacting phonon states weighted by the canonical density operator for an uncoupled phonon bath,\cite{Dunn1975,Mahan2000} namely
\begin{align}
\mathcal{G}\left(k,t\right) & =-i\Theta\left(t\right)\frac{\text{Tr}\left[e^{-\beta H_{p}}a_{k}(t)a_{k}^{\dagger}(0)\right]}{\text{Tr}\left[e^{-\beta H_{p}}\right]},
\nonumber \\ &
\equiv-i\Theta\left(t\right)\langle a_{k}(t)a_{k}^{\dagger}(0)\rangle.\label{eq:-1-1}
\end{align}
As will be useful in the next subsection, we also define the quantity
\begin{align}
\Phi\left(k,t\right) & \equiv\log\frac{\mathcal{G}\left(k,t\right)}{\mathcal{G}_{0}\left(k,t\right)},
\end{align}
where
\begin{align}
\mathcal{G}_{0}\left(k,t\right) & =-i\Theta\left(t\right)\langle e^{iH_{e}t}a_{k}e^{-iH_{e}t}a_{k}^{\dagger}\rangle.
\end{align}
Finally, most comparisons with exact calculations will be made via consideration of the spectral function, defined as
\begin{align}
A_k(\omega) & =-\frac{1}{\pi}\Im\left[\int_{-\infty}^{\infty}dte^{i\omega t}\mathcal{G}\left(k,t\right)\exp\left(-\gamma t\right)\right],
\label{eq:spectral_function_def}
\end{align}
which most closely connects the one-particle Green's function to angle resolved photoemission experiments.\cite{Mahan2000}
Here, $\gamma$ is a broadening parameter which is used to enable comparison with VD calculations, serves to dampen recurrences for calculations with a small number of sites, and which may be considered as an effective ``experimental" resolution for the spectral function itself.

\subsection{Cumulant Expansion for $\mathcal{G}\left(k,t\right)$\label{subsec:Cumulant-Expansion-for}}

Various perturbative approaches have been developed for the explicit calculation of Green's functions such as
$\mathcal{G}\left(k,t\right)$. The standard approach, which we shall not follow here, follows the now well-established rules of quantum field theory.\cite{Mahan2000,Fetter2003}
Instead, we will follow the ``linked-cluster'' or cumulant approach perhaps first used in the form we employ by Brout and Englert,\cite{PhysRev.120.1519, PhysRev.130.409} and applied to polaron models originally by Mahan, Dunn and others.
\cite{mahan1966phonon, Dunn1975}

First, consider the difference between expansions of the moment generating function (MGF) and the cumulant generating function (CGF). For a (classical) Gaussian
random variable $X$, an expansion of the MGF truncated at second-order will only approximately describe the MGF,
\begin{align}
\langle e^{-igX}\rangle & =1-ig\langle X\rangle-\frac{g^{2}}{2}\langle X^{2}\rangle+\mathcal{O}\left(g^{3}\right).
\end{align}
However, the CGF in this case is exactly described by a second-order expansion,
\begin{align}
\log\langle e^{-igX}\rangle & =-ig\langle X\rangle-\frac{g^{2}}{2}\left(\langle X^{2}\rangle-\langle X\rangle^{2}\right).
\end{align}
By taking the logarithm of the Gaussian MGF before expanding, one effectively resums an infinite number higher-order terms in the MGF. The inclusion of even approximate terms of higher-order in the perturbation expansion leads one to expect that a cumulant method can be accurate, especially if the expanded quantity is ``nearly Gaussian'' in the sense of having small cumulants of order higher than second.

The CE is readily adaptable for perturbative calculation of both
thermodynamics and quantum dynamics, where, like the MGF, both the Boltzmann factor and the propagator are exponential functions to be averaged, albeit in time-ordered form.\cite{van1992stochastic}
Analogous to
a MGF, the one-particle Green's function
\begin{align}
\mathcal{G}(k,t) & =\mathcal{G}_{0}\left(k,t\right)\biggr<e_{T}^{-i\int_{0}^{t}d\tau\hat{V}\left(\tau\right)}\biggr>_{k}\\
\mathcal{G}_{0}\left(k,t\right) & =-i\Theta(t)e^{-i\varepsilon_{k}t},
\end{align}
can also be calculated approximately via a perturbative calculation
of $\biggr<e_{T}^{-i\int_{0}^{t}d\tau\hat{V}\left(\tau\right)}\biggr>_{k}$
in powers of a coupling constant. Here, the $k$ subscript denotes
the average over all one-electron states with electronic momentum
$k$, the $T$ subscript denotes time-ordering, and the hat designates
$\hat{V}\left(\tau\right)$ as an operator in the interaction picture. 

The $\mathcal{M}^{\text{th}}$-order CE (linked-cluster) for
$\mathcal{G}\left(k,t\right)$ is given by
\begin{align}
\mathcal{G}_{\mathcal{M}}(k,t) & =\mathcal{G}_{0}\left(k,t\right)\exp\left[\Phi_\mathcal{M}(k,t)\right],\label{eq:-3.5}
\end{align}
and likewise $A_{\mathcal{M}}\left(k,\omega\right)$ is calculated
via the Fourier transform of $\mathcal{G}_{\mathcal{M}}\left(k,t\right)$. 
Here, $\Phi_{\mathcal{M}}(k,t)$ is the sum of the cumulants $C_{\mu}$ up 
to order $\mathcal{M}$.
The procedure for constructing the cumulants $C_{\mu}\left(k,t\right)$
from the moments $M_{\mu}\left(k,t\right)$ is well-known.\cite{Mahan2000}
For models of the form given in Eqns. (\ref{eq:1}-\ref{eq:-3.7}), the first few cumulants (up to the fourth-order cumulant) are explicitly given by
\begin{align}
C_{1} & =0,\\
C_{2} & =e^{i\varepsilon_{k}t}M_{2},\\
C_{3} & =0,\\
C_{4} & =e^{i\varepsilon_{k}t}M_{4}-\frac{1}{2}C_{2}^{2},
\label{eq:C4_general}
\end{align}
where
\begin{align}
M_{\mu}(k,t) & =\frac{(-i)^{\mu}}{\mu!}\int_{0}^{t}dt_{1}...\int_{0}^{t}dt_{\mu}\nonumber \\
&\times
\biggr<T\left\{ \hat{a}_{k}(t)\hat{V}(t_{1})...\hat{V}(t_{\mu})\hat{a}_{k}^{\dagger}(0)\right\} \biggr>.\label{eq:-3.4}
\end{align}
Here $T\left\{ \dots\right\} $ is the time-ordering operator which places later times to the left.

Calculation of the second-order and fourth-order CE for the Holstein
model requires computing $M_{2}$ and $M_{4}$. These moments
depend on $\varepsilon_{k}$ and the form of the EPI vertex, which
is a momentum-independent constant for the Holstein model. Expressions
for $M_{2}$ and $M_{4}$ where we evaluate the time integrals and leave
the momentum sums explicit are given in Appendix \ref{Appendix3}. 

\begin{figure*}[tb]
\begin{centering}
\includegraphics{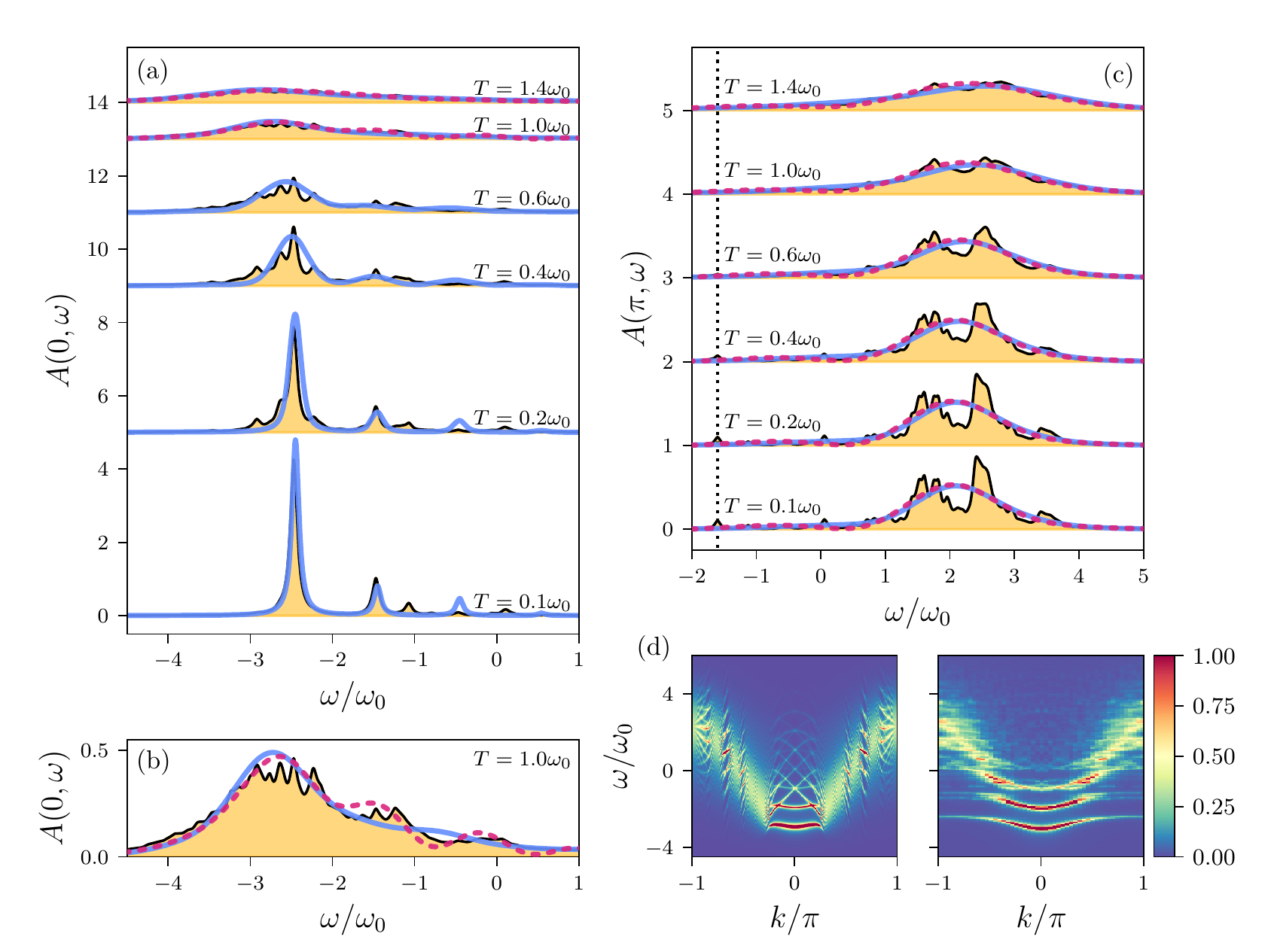}
\par\end{centering}
\caption{Spectral functions $A\left(\omega\right)$ for the 6-site Holstein
model calculated via the second-order CE (blue solid line), fourth-order CE (pink dashed line), and VD (thin black line and gold shading).
Model parameters: $\omega_{0}=t_0=g=1$. We use $\gamma=0.05$.
VD results are the same as those presented in from Fig. 1(a) of Bon{\v c}a \textit{et al}. \cite{Bonca2019}
(a) $k=0$ for a range of temperatures.
The fourth-order CE is only presented for the two highest temperatures because it is divergent at lower temperatures.
Both orders of the CE capture the most prominent structures of the VD result.
(b) $k=0$ at only $T=\omega_0$.
When the fourth-order CE is convergent it slightly corrects the quasiparticle energy and adds additional structure to the peak which better approximates the VD result.
(c)  $k=\pi$ for a range of temperatures.
The VD result demonstrates that there is significant structure 
in the spectral function
and a quasiparticle peak at
$\sim -1.5\omega_0$ (dotted black line) while the second- and fourth-order 
CE only broadly model the structured features in the spectral function.
In the fourth-order CE there is 
an addition peak centered around $- \omega_0$.
(d) Heat maps for second-order CE (left) and VD (right, from Ref. \citenum{Bonca2019}) at all momenta for $\omega_{0}=t_0=1$, $T=0.1$ and $g^2=2$. Note how in the CE the extra bright peaks at $k=0$ disperse into a series of shifted peaks which eventually coalesce into the incoherent polaron peak at $k=\pm\pi$. In contrast, the VD bands extend all the way from $k=0$ to $k=\pm\pi$.
}
\label{fig:six_site_zero_and_pi}
\end{figure*}

\subsection{Convergence of Cumulant Expansion\label{subsec:Convergence-of-Cumulant}}

Let us now examine some aspects of the convergence of the CE for the Holstein model as
a function of temperature and EPI coupling strength. Following the definition
of the CE in Eq. \ref{eq:-3.5}, a sufficient condition for the break down of the expansion occurs when successive higher-order cumulants $C_{\mu}$ are not relatively small. Therefore we examine the magnitude $C_{\mu}$. 

For $T \rightarrow 0$, the phonon occupation numbers $N_{0}$ vanish,
such that we may order the terms in Appendix \ref{Appendix3} as functions of the coupling strength for the finite, even cumulants
\begin{align}
C_{2n} & \sim g^{2n}.
\end{align}
The high temperature limit is slightly more subtle. Before performing
the time integrations in Eq. (\ref{eq:-3.4}), contracting
the phonon operators yields
\begin{align}
C_{2n} & \sim g^{2n}\prod_{i=1}^{n}\left[\coth\left(\frac{\beta\omega_{0}}{2}\right)\cos\left(\omega_{0}\tau_{i}\right)-i\sin\left(\omega_{0}\tau_{i}\right)\right].
\end{align}
Taking the high temperature limit,
\begin{align}
C_{2n} & \sim g^{2n}\prod_{i=1}^{n}\frac{2}{\beta\omega_{0}}\cos\left(\omega_{0}\tau_{i}\right)\sim\left(\frac{2g^{2}}{\beta\omega_{0}}\right)^{n}.
\end{align}
Thus, we see from these two limits that the CE in the time domain
breaks down for large $T$ and $g$ such that, schematically, the expansion
is governed by
\begin{equation}
\sim \max\left[g^{2},\frac{2g^{2}}{\beta\omega_{0}}\right].
\end{equation}

According to Dunn, \cite{Dunn1975} the CE should also give a reasonable
description of $A\left(k,\omega\right)$ at high enough temperatures
and strong enough coupling such that ($N_{0} =\frac{1}{e^{\beta\omega_{0}}-1}$)
\begin{align}
g N_{0} & \gtrsim1,
\end{align}
and/or 
\begin{align}
g\left(N_{0}+1\right) & \gtrsim2,
\end{align}
which supersedes the condition for convergence given above.
In this regime the long time behavior of $\mathcal{G}\left(k,t\right)$ is quickly
damped and $A\left(k,\omega\right)$ is broadened to such a 
degree to as to wash out all sharp spectral features. We shall see evidence of these
behaviors in the following sections.

\section{Results\label{sec:Results}}

\begin{figure}[tb]
    \centering
    \includegraphics{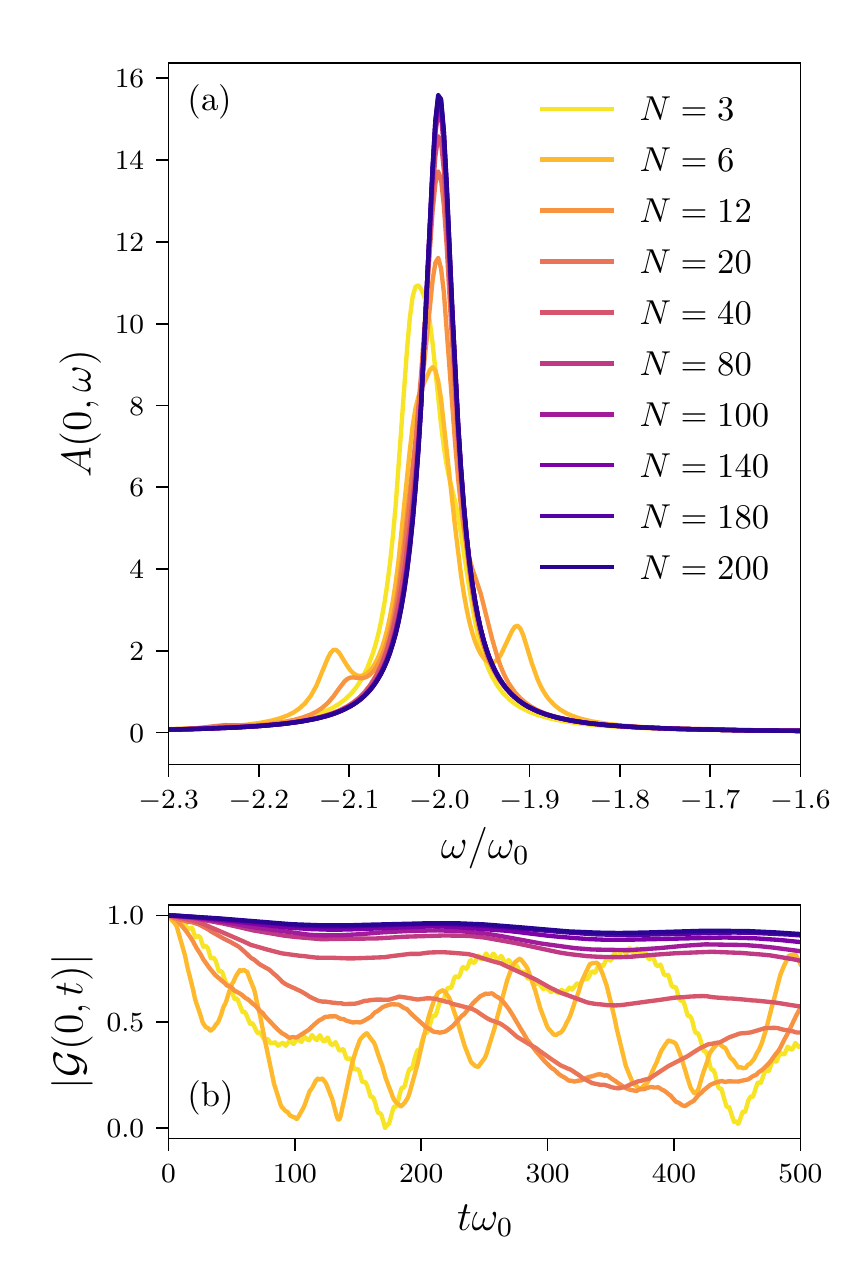}
    \caption{One-phonon exact diagonalization approximation results of Holstein model at increasing system sizes. $\omega_0 = t_0 = T = 1; g=0.25$
    (a) Spectral function as a function of system size demonstrates disappearing fine structure (b) Magnitude of the Green's function 
    in time as a function of system size demonstrates disappearance of sharp beats.}
    \label{fig:k_phonon}
\end{figure}

\begin{figure*}[tbh]
    \centering
    \includegraphics{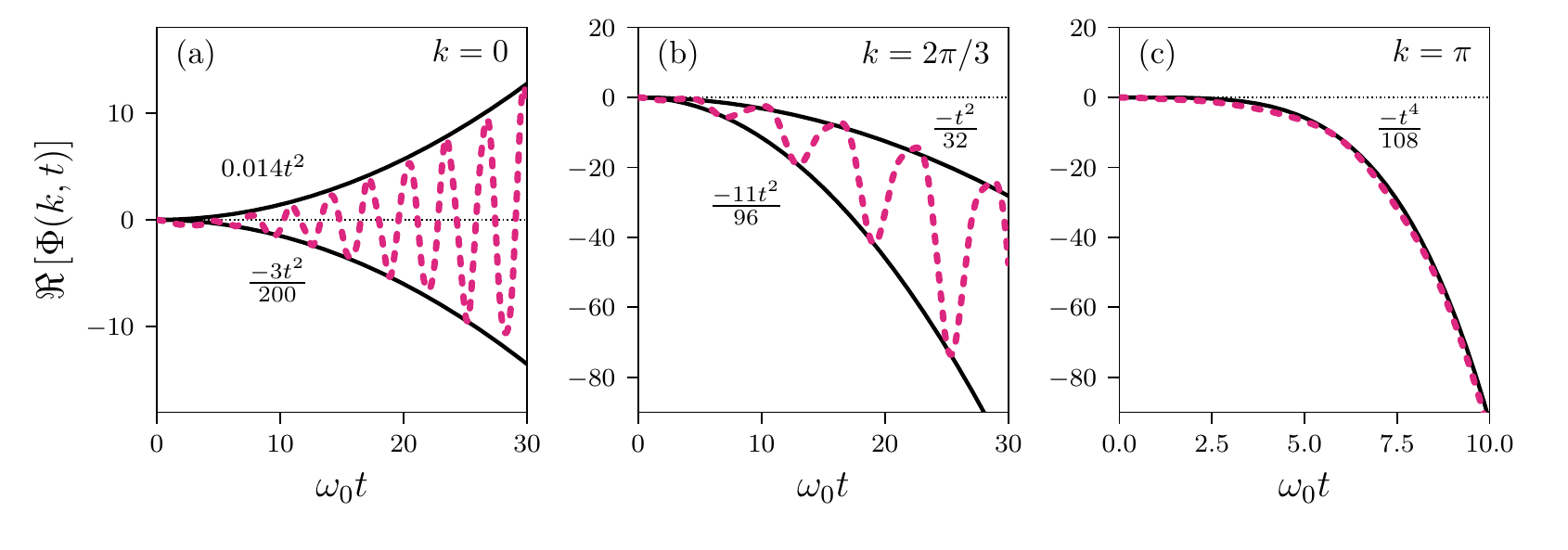}
    \caption{The fourth-order CE numerical results (dashed red) for $N=6$ at $g=\omega_0=t_0=1$ and $T=0$ compared with the leading order envelope functions (solid black) at (a) $k=0$, (b) $k=2\pi/3$, and (c) $k=\pi$. The functional form of the envelope is given in each figure and describes the general shape of the numerical results.
    A discussion of the origin of these envelope functions is found in the main text and Appendix A.}
    \label{fig:divergence_six_site_example}
\end{figure*}

Recently, Bon{\v c}a \textit{et al.}~published the first exact 
temperature-dependent spectral function for the single-particle Holstein
model using VD.\cite{Bonca2019}
Since the VD and finite temperature Lanczos methods are well detailed in the literature, we do not review them here.\cite{Bonca2019,Trugman1990,Bonca1999,Ku2002,Jaklic2000,Prelovsek2013}
Due to the expense of the approach, the finite-temperature calculations of Bon{\v c}a \textit{et al.} were limited to small system sizes of 6 and 12 sites.  It should be noted that at the level of heat maps of the k-dependent spectral function, the 6 and 12 site results differ by only a small amount. On closer inspection, small finite size effects are apparent, as will be discussed below.  Recently, new techniques have been developed that are capable of providing exact finite temperature spectra in the single-particle polaron models for larger systems.
\cite{Jansen2020}
However comprehensive results for larger lattices in the Holstein model have not yet been published, and thus we compare only to the work of Ref.\citenum{Bonca2019}.

To judge the accuracy of the CE and compare its performance to the exact results of Bon{\v c}a \textit{et al.}, \cite{Bonca2019} we now consider the CE approximation for spectral functions for a 6-site Holstein model at the band bottom ($k=0$) and the band edge ($k=\pi$).  We restrict the comparison to the intermediate coupling regime where $g=t=\omega_{0}$. In Fig. \ref{fig:six_site_zero_and_pi} we plot the $k=0$ and $k=\pi$ second- and fourth-order CE spectral functions along with data from Ref. \citenum{Bonca2019}. For both momenta, the second-order CE captures the broad structure of the spectral function reasonably well, and for $k=0$ the results are quantitatively accurate at both $T=0$ and at high temperatures $T \ge \omega_{0}$.  In particular, the second-order CE captures the quasi-particle peak and the first vibronic satellite peak at $T=0$ in excellent agreement with VD.  At higher temperatures, the central features of these peaks are well captured, however the fine structure superposed on the quasi-particle peak exhibited by the exact VD spectra is absent in the CE spectra.  We will see below that this fine structure is a consequence of the small lattice size, and thus the CE approximation does not properly capture this type of finite lattice effect.

At the band edge ($k = \pi$) the results produced by the second-order CE are not as encouraging, as illustrated in in Fig. \ref{fig:six_site_zero_and_pi}(c).  The VD data has two important features: a quasi-particle peak at low energy ($\sim - 1.5\omega_0$), and a broad vibronic wing with a split peak structure centered around ~$2 \omega_0$.  The second-order CE misses the peak structure of the exact spectral function entirely, and instead can be described as a single broad peak centered near the average value of the peak intensity found in the exact VD result.  Again, as temperature increases and the features of the spectral function broaden, the CE result becomes more and more accurate, reflecting the fact that the CE properly accounts for the spectral bandwidth even for $k=\pi$.  The fact that the CE is accurate away from $k=0$ for temperatures $T \ge \omega_{0}$ has important practical implications for the use of the CE to study transport phenomena, a topic we will return to before concluding.

The difference in accuracy of the CE between the $k=0$ and the $k=\pi$ cases is seen generally across the full range of wave vectors.  More specifically, we find that the $k=0$ case is the only case for which the CE is in quantitative agreement with exact VD results for low temperatures.  A full comparison of the exact and approximate CE spectral functions across the entire band can be found in Fig. \ref{fig:six_site_zero_and_pi}(d).  Here, several features are notable.  The fact that for $k=0$ the CE predicts a prominent series of small peaks beyond the first satellite spaced by ~$\omega_{0}$, in reasonable agreement with the exact VD results, is actually the result of an incorrect intensity crossing structure which renders the satellite behavior for all $k \neq 0$ inaccurate.  As we will discuss below, this behavior is the result of the manner in which the CE approximates higher-order multi-phonon scattering terms.  Note as well that for $k \neq 0$ there is fine structure in the high intensity band.  This behavior is a finite size effect of the CE which vanishes when the number of lattice sites tends to infinity, as we will discuss later in this work.  Such finite size effects are distinct from the true finite sized behavior exhibited in the VD results discussed above, and do not reflect the correct formation of structure exhibited in the satellite region of the exact spectra.

We next turn to a discussion of the corrections to the second-order CE provided by the fourth-order CE. In general, when the fourth-order CE is well-defined for the parameter regime of the Holstein model studied here, it only subtly alters the behavior found from the second CE.  In Fig.~\ref{fig:six_site_zero_and_pi}(b) we provide a close-up of the $k=0$, $T=\omega_{0}$ case found in Fig.~\ref{fig:six_site_zero_and_pi}(a).  It can be observed that in general the fourth-order CE indeed redistributes spectral weight correctly, with the exception of a small region of negative spectral weight for $\omega > 0$.  The fact that the fourth-order CE does not guarantee positivity of the spectra has been discussed in several previous works.\cite{Gunnarsson1994, Zuehlsdorff2019, Anda2016}
Gunnarsson {\em et al.} attribute this problem to the particular analytical form of the terms retained at fourth-order in the CE.  More problematic is the fact that for some parameter regimes the fourth-order CE is not well defined due to unbounded growth in the time domain of some of the terms in the expansion.
\cite{Gunnarsson1994}
We will see below that these terms take a similar form to those pointed out as contributing to negative spectral weight by Gunnarsson {\em et al.}  Thus, these two issues appear to be connected.  In Fig.~\ref{fig:six_site_zero_and_pi}(a) fourth-order CE results are not shown for $k=0$ and $T \leq 0.6\omega_{0}$ due to the divergence in the time domain of the fourth cumulant.
In the next three subsections we will investigate more deeply
several of the features exposed here for the finite-sized Holstein chain before turning to the CE in the thermodynamic limit.

\begin{figure}[tbh]
\begin{centering}
\includegraphics{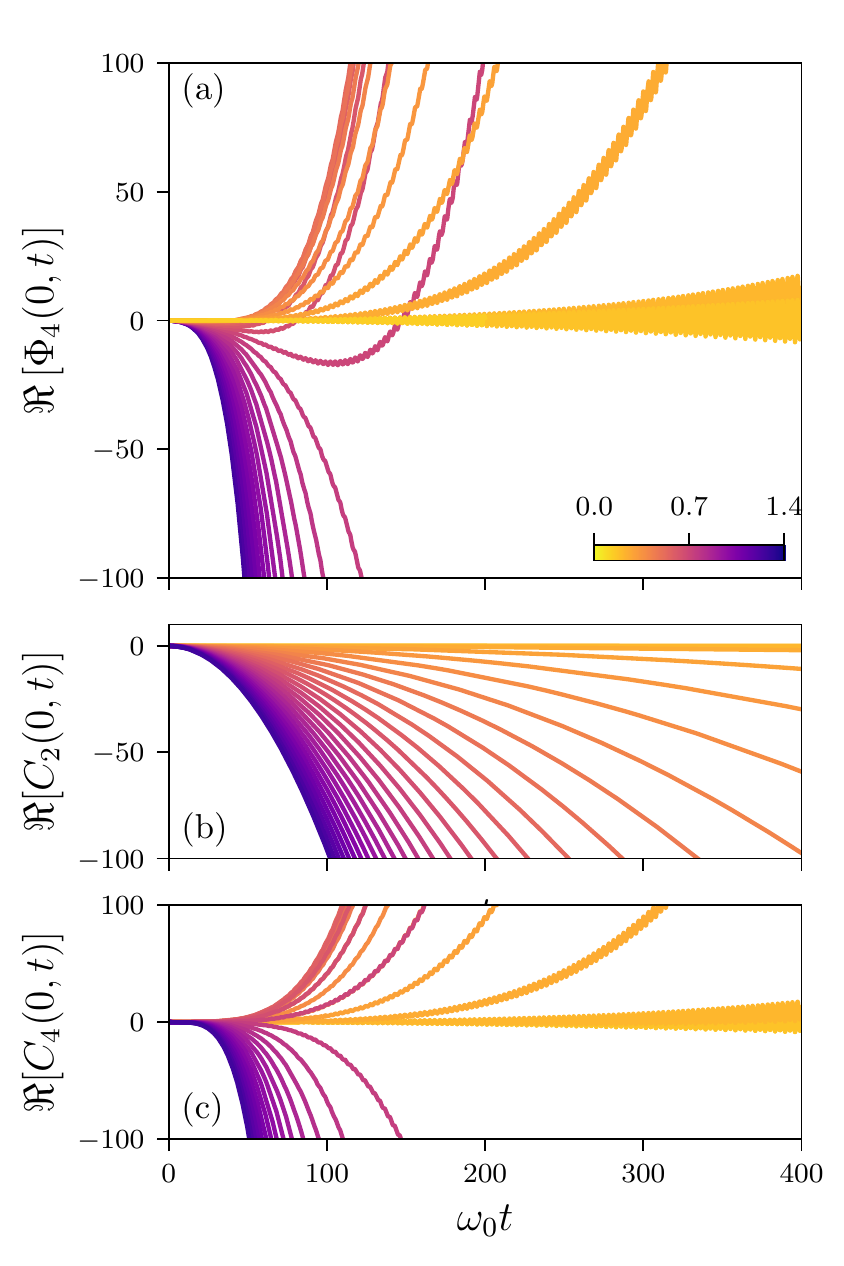}
\par\end{centering}
\caption{Breakdown
of the time dependence of (a) $\Phi_4(0,t)$  and its dependence on (b) $C_2(0, t)$ and (c) $C_4(0,t)$,
highlighting the temperature driven transition from divergent to non-divergent $\mathcal{G}(0,t)$.
Here, $\omega_0 = t_0 = 1, g=0.25$, $k=0$ and $T = [0, 1.4] $.
$\Phi_4(0,t)$ transitions from divergent to non-divergent between $T=0.7$ and $T=0.75$, and
because $C_2(0,t)$ does not predict a divergent $\mathcal{G}(0,t)$ for any temperature, the transition from divergence or non-divergence is dictated by the transition in $C_4(0,t)$.}
\label{fig:fourth_order_divergence}
\end{figure}

\subsection{Finite Size Effects\label{sec:cumulant_finite_size}}
In the discussion of results for the 6-site lattice presented above, we mentioned
several aspects of both the exact VD results as well as the results of the CE that warrant further discussion.
In this subsection we focus on one such feature, namely the role played by the small lattice size, and the implications for the failure of the CE to capture these effects.  In particular, we focus now on the small ripples that appear in the main quasi-particle region of the VD spectra for a 6-site system in the regime $0.1\omega_{0} \geq T \geq 0.6\omega_{0}$.  We explicitly demonstrate that these features are due to the small lattice size, and thus the failure of the CE to capture this type of finite size effect is not relevant in the thermodynamic limit.  Indeed, the expected change in the spectral function in transitioning from small finite size systems to the $N \rightarrow \infty$ limit plotted in the manner of Fig. 1(e) will largely appear confined to smoothing the intensity modulation of the most prominent spectral features.

To shed light on the type of finite size effects expected to arise in small lattice systems, and to reveal why these effects show up prominently only at low to intermediate values of the temperature, we turn to exact diagonalization for finite sized systems in the one-phonon sector.
This approach is outlined in Appendix C. Due to the strong restriction on the phonon excitations allowed,  we do not aim for quantitative results and merely expose the qualitative nature of the spectral features associated with the quasi-particle peak as the system is tuned from finite to infinite lattice size.

In Fig.~\ref{fig:k_phonon} we show the behavior of the spectral function and the real-time behavior of the one-particle Green's function for $k=0$ for a weakly coupled electron-phonon system $(g=0.25)$ with parameters $\omega_{0}=t=T=1$.  The behavior of the Green's function in the time domain reveals the existence of higher frequency beating behavior superposed on lower frequency oscillations.  The high frequency behavior is related to recurrences due to transitions associated with the discreteness of the spectrum in the small $N$ limit.  Such behavior will manifest most strongly at intermediate temperatures, where thermally-populated low-lying states can participate in producing the observed beating behavior but where the temperature is not so high that damping effects dominate the decay of the Green's function. We note that already by $N=20$ the erratic high frequency behavior vanishes, although finite size effects are still present.  In the frequency domain, spectral functions of finite size systems with $N \leq 12$ exhibit small secondary peaks similar to the behavior exhibited in Fig. 1(a).

The finite size behavior and the inability of the CE to capture it is similar to that seen in purely electronic systems.  In particular, McClain {\em et al.} have studied the spectral function of the electron gas with coupled-cluster and cumulant-based techniques in finite sized systems.\cite{McClain2016}  Here the CE also shows a relative inability to reproduce structure associated with the discrete nature of finite sized systems.  We emphasize that the structure of the spectral function seen in small systems in the Holstein model discussed in this section are distinct from larger scale features for $ k \neq 0$ such as that seen in Fig. 1(c) which are also absent in low-order CE calculations.  The more important failure to reproduce these larger scale features is expected to persist in the $N \rightarrow \infty$ limit.

\subsection{Divergences in the fourth-order CE\label{subsubsec:Divergences-in-the}}

\begin{table}[tb]
    \centering
    \begin{ruledtabular}
    \begin{tabular}{c c c l}
         $N$      &  $g$     & $T/\omega_0$    & $\omega_0 t_{\Re[\Phi_4] > 0}$\\
         \hline\hline
        6  &  1.00  &  0.40 & 7.3 \\
        6  &  1.00  &  0.60 & 10.7 \\
        6  &  1.00  &  0.70 & 23.2 \\
        6  &  1.00  &  0.72 & 80.0 \\
        6  &  1.00  &  0.73 & $> 10^4$ \\
        \hline
        6  &  0.75  &  0.00 & 7.3 \\
        6  &  0.50  &  0.00 & 10.7 \\
        6  &  0.25  &  0.00 & 20.2 \\
        6  &  0.10  &  0.00 & 51.6 \\
        \hline
        6  &  1.00  &  0.00 & 7.1 \\
        12  &  1.00  &  0.00 & 10.2 \\
        50  &  1.00  &  0.00 & 24.9 \\
        100  &  1.00  &  0.00 & 49.0 \\
        150  &  1.00  &  0.00 & 73.3 
    \end{tabular}
    \end{ruledtabular}
    \caption{ $t_{\Re[\Phi_4] > 0}$ for varying system sizes, coupling strengths and temperatures.
    Increasing the system size or decreasing the coupling can push the onset of the 
    divergence to longer times. Raising the temperature past some
    transition temperature fully removes the divergence. Model parameters: $t_0=\omega_0=1; k=0$
    }
    \label{tab:divergence_time}
\end{table}

Fig. 1(b) illustrates that, aside from the unphysical appearance of regions with a (small) negative spectral weight for $\omega \gtrsim 1.1\omega_0$ (not shown), the fourth-order CE improves upon the second-order
CE for the spectral function at $k=0$ at higher temperatures. 
However, as mentioned above, for the same $k$ value at low temperatures, the fourth-order CE is divergent at longer times, and thus truncated higher-order CEs cannot always be used to systematically improve upon low-order results. Here we focus on the factors which can shift the fourth-order CE between well-behaved and divergent at long times to better understand where corrections to the second-order CE are applicable. 
We will demonstrate that in the Holstein model, the divergence of the fourth-order CE depends intimately on the wave vector, system size and temperature under consideration, and is closely connected to the issue of negative spectral weight first pointed out for this model by Gunnarsson {\em et al.}~\cite{Gunnarsson1994}

The CE is an exponential function of the quantity $\Phi(k,t)$ defined in Sec.~1B for which physical results require $\Re\left[\Phi(k,t)\right] \le 0$ for all times. In addition, at finite temperatures the requirement
$\lim\limits_{t\to\infty} \Re\left[\Phi(k,t)\right] \to -\infty$ must hold, reflecting the finite lifetime of quasi-particles. It is easily checked that the second-order CE always satisfies these requirements. In particular, 
$-g^2 t^2 (2 N_0 + 1) \leq \Re[\Phi_2(k,t)] \leq 0,$ and thus
the second-order CE never diverges.

With these considerations in hand, we focus on the fourth-order CE, characterizing the divergence of the fourth-order term $\Phi_{4}(k,t)$ by the quantity $t_{\Re[\Phi_4] >0}$, which marks the earliest time where $\Re[\Phi_4] >0$.
In Table \ref{tab:divergence_time} we compile $t_{\Re[\Phi_4] >0}$ for the Holstein model with the same parameters as found in Fig. 1 for the spectral function at $k=0$ as a function of temperature and the number of lattice sites.
Several aspects of the data are worthy of note.  We focus first on the fact that as $T$ approaches a temperature between $T=0.72$ and $T=0.73$, the divergence is abruptly pushed from a finite time to infinite time for all practical purposes. This behavior is consistent with the results plotted in Fig. \ref{fig:six_site_zero_and_pi}(a), 
where only the cases $T=1.0\omega_0$ and $T=1.4\omega_0$ have non-divergent fourth-order CE results. 

The root of this abrupt behavioral change in the long-time limit of $\Phi_{4}(k,t)$ becomes manifest upon examining the analytical forms of the individual cumulant terms. Details may be found in Appendix \ref{Appendix3}. 
Direct examination of $\Phi_4(k, t)$ at zero temperature reveals terms of the form $t e^{i \alpha t}$ where $\alpha$ is some real number. 
Care must be taken with the evaluation of $C_4(k, t)$, as it contains many apparent singularities which are actually well-defined when appropriate limits are taken.
Depending on the particular limit and the values of $t_0$ and $\omega_0$, $C_4(k, t)$ contains real-valued terms 
which may diverge linearly, quadratically, or quartically in time.
Some of the seemingly divergent terms of order $t^2$ in $M_4(k,t)$ are exactly cancelled by the transformation from moments to cumulants in Eq. \ref{eq:C4_general}.

\begin{figure}[tb]
\begin{centering}
\includegraphics{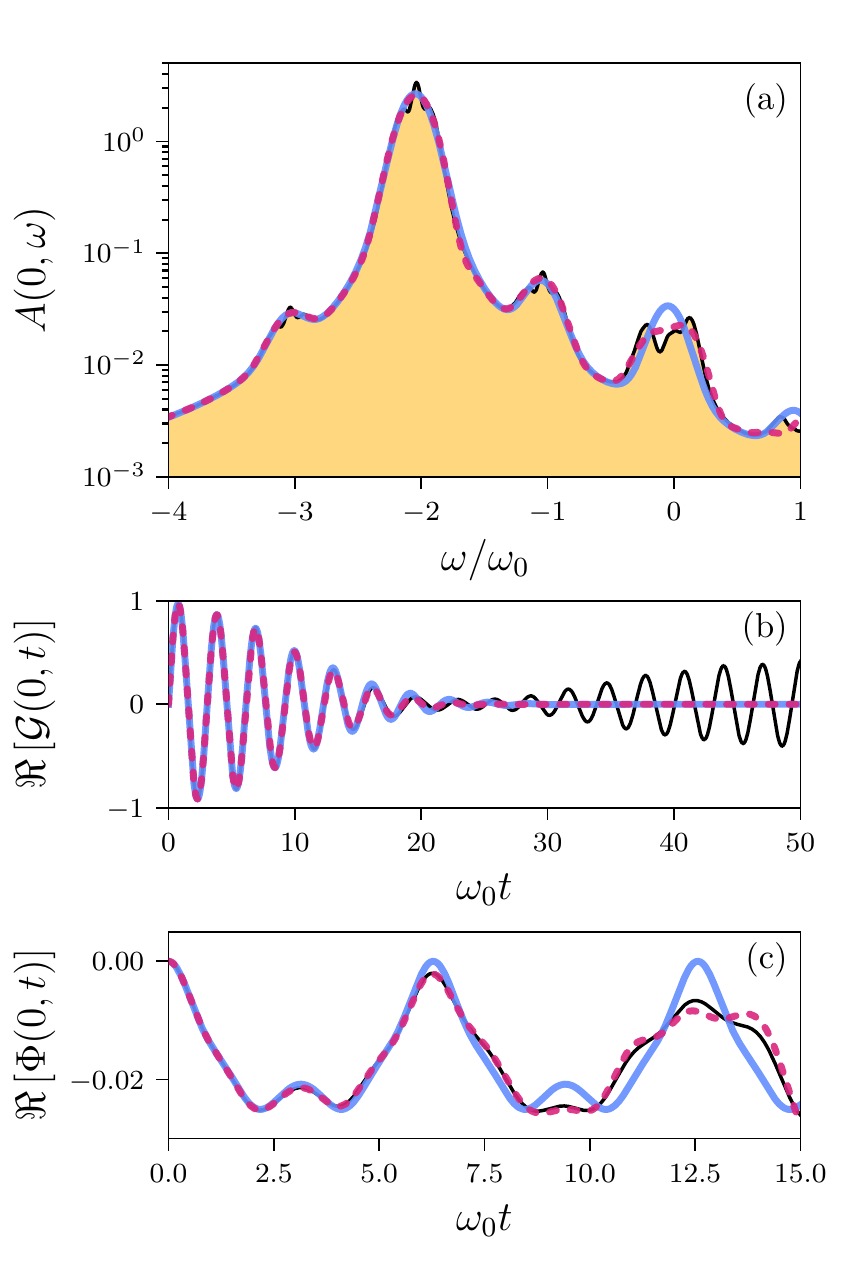}
\par\end{centering}
\caption{
(a): Comparison of $A_2$ (blue solid line), $A_4$ (pink dashed line)
and $A_{\infty/\text{HEOM}}$ (thin black line and gold fill) for $T=1.0\omega_0$. 
(b): Comparison of the real parts of $\mathcal{G}_{2}\left(t\right)$ and $\mathcal{G}_{4}\left(t\right)$ with
$\mathcal{G}_{\infty}\left(t\right)$ for $T=1.0\omega_0.$
(c): Comparison of $\Phi_{2}\left(t\right)$ and $\Phi_{4}\left(t\right)$
with $\Phi_{\infty}\left(t\right)$ for $T=0.0\omega_0.$
 Model parameters: $N=6; k=0;\omega_{0}=t_0=1$,
$g=0.25$, $\gamma = 0.04$ (spectral function only).
HEOM calculations are performed using modified versions
of \textit{PHI }\cite{Strumpfer2012} 
and \textit{pyrho}.\cite{Berkelbach}
}
\label{fig:heom_ce_small_g}
\end{figure}

Here, we present the leading-order contributions to the fourth cumulant for $N=6, t_0 = \omega_0, T=0$ and $k=0$,
\begin{equation}
    \Phi_4(0, t)= \frac{-g^4t^2\left(2 e^{-5 i t \omega_0}+25 e^{-2 i t \omega_0}\right)}{1800 \omega_0^2} + o(t^2).
\label{eq:divergent_term_k_zero}
\end{equation}
Since the exponential functions in Eq.~\ref{eq:divergent_term_k_zero} contain no real damping, 
$\Phi_4(0,t)$ diverges quadratically in time.
The envelope growth rapidly becomes the only significant term in the expansion.
This is illustrated in Fig.~\ref{fig:divergence_six_site_example}(a) where the case of $g=\omega_0=1$ is explicitly shown, and $\Re[\Phi(0, t)]$ indeed grows along the upper and lower bounds of Eq. \ref{eq:divergent_term_k_zero}. 
Although the case presented here is quadratically divergent, this is specific to $t_0=\omega_0$.
A more general version of Eq.~\ref{eq:divergent_term_k_zero} contains
only linear divergences (albeit many of them), while the proper evaluation of the limit $t_0 \to
\omega_0$ introduces quadratically growing terms.
It is worth noting that terms of this general form were also found by Gunnarsson
\textit{et al.} in their zero-temperature CE study of similar polaron models.\cite{Gunnarsson1994} These authors determined that such terms give rise problematic negative spectral weight, however they did not report a divergent behavior in $G\left(k,t\right)$.
We will return to the issue of negative spectral weight below.

Inspection of Eq.~\ref{eq:divergent_term_k_zero} also makes clear why decreasing the value of the electron-phonon coupling delays the onset of the divergence, as seen in Table 1.  
The divergent portion of the fourth-order cumulant is scaled by
$g^4$, so it is quite expected that smaller $g$ decreases the time scale of divergence. This is demonstrated in Table \ref{tab:divergence_time} with the example of a six site system where decreasing the value of $g$ indeed increases the divergence time. 

In a similar vein, we can examine the terms in $\Phi_4(\pi, t)$ and $\Phi_4(2\pi/3, t)$ to understand why, for $N=6$ and $\omega_0=t_0$, the fourth-order CE is non-divergent at zero temperature. 
The case of $k=\frac{2\pi}{3}$ is very similar in form to $k=0$ but critically
contains an extra constant in the expression for the leading term, namely
\begin{equation}
    \Phi_4\left(\frac{2\pi}{3}, t\right)=\frac{-g^4t^2 \left(4 e^{i t \omega_0 }+e^{-2 i t \omega }+6\right)}{96 \omega_0 ^2} + o(t^2).
    \label{eq:phi_4_2pid3_leading}
\end{equation}
Because of the $-6t^2g^4/96\omega_0^2$ term in Eq.~\ref{eq:phi_4_2pid3_leading}, $\Re[\Phi_4(2\pi/3, t)]$ tends towards negative infinity quadratically, which
corresponds to a strongly damped $\mathcal{G}(2\pi/3, t)$.
This is shown in Fig. \ref{fig:divergence_six_site_example}(b) for the case of $g=\omega_0=t_0=1$.
As in the case of $k=0$, the avoidance of an unphysical divergence is
specific to the choice that $t_0=\omega_0$, so it merely serves as a demonstration of one way in which the terms in the fourth-order CE conspire to avoid divergences issues for a specific set of parameters at a specific wave vector.

While these examples demonstrate how terms quadratic in time in
the expansion of $\Phi_4(k,t)$ can
lead to a convergent or divergent approximation to $\mathcal{G}(k,t)$, leading
terms of even higher-order in time are possible.
An example of this occurs at $k=\pi$, where in Fig. \ref{fig:six_site_zero_and_pi}(c) we
observe that the fourth-order CE is always well behaved. 
Once again taking $N=6, \omega_0=t_0$ and $T=0$, we can show that the leading-order 
divergence for $k=\pi$ is
\begin{equation}
    \Phi_4(\pi, t) = \frac{-g^4 t^4}{108}  + o(t^4).
\end{equation}
Unlike the behaviour expressed by Eq. \ref{eq:divergent_term_k_zero}, the leading-order divergence of $\Phi_4(\pi, t)$ is not oscillatory;
rather, it is strictly negative and quartic, rendering irrelevant any oscillating and increasing terms of slower growth.
This qualitative difference in behavior between the $k=0$ and $k=\pi$ cases ensures a non-divergent Green's function at the band edge even at low temperatures.

The divergent terms discussed above are related to the double pole structure described by Gunnarsson {\em et al}.\cite{Gunnarsson1994}  Analyzing the behavior of the fourth cumulant in the short-time limit, a function of the form $\exp\left( \alpha t^n e^{-z t} + \dots \right)$ can be linearized to give $1 + \alpha t^n e^{-z t} + \dots$. Here, $\alpha$ is a complex coefficient, $n$ is a positive integer, and $z$ is purely imaginary.
The Fourier transform of a function of this form will be proportional to the $n^{\text{th}}$ derivative of a delta function centered at $z$, and this feature will be present in the spectra even if the overall CE is convergent. While this argument is approximate, as it relies on the short-time dynamics, it nonetheless makes clear the connection between negative spectral weight and the potential for divergent behavior in the fourth- (and presumably higher)-order CE.  Since the CE to all orders is exact, the cancellation of these problematic terms at high-orders must occur, albeit clearly in a complicated manner which likely obviates the possibility of removing such terms in lower-order versions of the the expansion in a reliable way.
In most applications we are interested in the $N \rightarrow \infty$ limit, 
and here, as shown in Sec. \ref{sec:thermodynamic_limit}, we note that for some wave-vectors divergences are suppressed with increased system size. 
As empirically demonstrated in Table \ref{tab:divergence_time} for $k=0$, 
the onset time of the divergence grows linearly with the system size.
To understand this behavior, we again consider which terms are present in the summation
of the expressions for the fourth-order cumulant.
As the system size increases, the number of terms in the momentum sums over $q_1$ and $q_2$ grows as $N^2$, while
the weight of each individual term decreases in magnitude as $N^{-2}$.
Singularities in $\Phi_4(k,t)$ that produce quadratic growth in time only occur when
specific energetic conditions are met.
A few examples of these conditions are 
$\epsilon_{k+q_1} - \epsilon_k + \omega_0=0$,
$\epsilon_{k+q_1} - \epsilon_{k+q_1+q_2} + \omega_0=0$,
and 
$\epsilon_{k+q_1+q_2} - \epsilon_{k} + 2\omega_0$=0.
Crucially, these conditions exist only on one-dimensional lines 
in the space of $q_1$ and $q_2$.
Thus, the ratio of the non-singular evaluations to the total number of evaluations falls of at least as $1/N$, as we observe numerically. 
In particular, for $k=0$ the non-singular term occurs in $61.1\%$ of evaluations of the momentum sum for $N=6$, $99.7\%$ of the evaluations for $N=600$, and $99.8\%$ of the evaluations for $N=1200$.
Thus, for very large system sizes we can drop all of the singular cases of $q_1$ and $q_2$ by recognizing that
the ratio of singular cases to non-singular disappears as $\sim 1/N$. 
The true thermodynamic limit of $C_4(k,t)$ corresponds to a principle value integral
over momentum space with real terms at most linearly divergent in $t$.

While the preceding argument justifies why the non-linear in time divergences present in small systems disappear as $N\to\infty$ for $k=0$, it does not explain
why linear time divergences do not appear.
We now heuristically argue that a distinct type of behaviour suppresses divergent growth in time as $N\to\infty$ for some wave vectors.

With the remaining linear terms proportional to $t e^{i\alpha t}$ where $\alpha \in \Re$, 
the momentum sum in the fourth-order cumulant
becomes one of many oscillating exponential functions,
each with weight $N^{-2}$.
As the frequencies in the exponential become continuously distributed, 
interference of the many out of phase
components can delay the onset of divergence to arbitrarily long times.
It must be noted that this cancellation depends on specific properties of the
unperturbed energy dispersion which are not trivially satisfied at all $k$.
Nonetheless, we find numerically that as 
$N$ tends towards an infinite number of sites for both $k=0$ and $k=\pi$, the first constructive beat
is pushed to $t=\infty$, hence
the results in section \ref{sec:thermodynamic_limit}
are well-behaved at all temperatures for those values of $k$ unlike for the case $N=6$.
We have not been able to uncover a deeper analytical argument for this behaviour,
and must appeal to numerical heuristics, which are presented in detail in Sec. \ref{sec:thermodynamic_limit}.

Since the infinite-order CE provides an exact representation of the dynamics, it must be true that even higher-order cumulant terms eventually
conspire to remove the 
divergent terms at lower-orders.
However, because $C_n(k,t) \propto g^{n}$, we know that the higher-order terms can not directly cancel the lower-order divergences, and instead must form the series representation of a well-behaved exact $\Phi(k,t)$.
There are a number of approaches one could attempt to remove these divergences, but all presume some knowledge of the higher-order terms in the CE.
For an approximate means of resumming higher-order cumulants, we refer the reader to 
the self-consistent cumulant approximation in the companion paper.\cite{RobinsonCumulantII}

\subsection{Convergence to Exact Result: Short-Time Analysis\label{sec:time_convergence}}
The results in the previous subsection illustrate that the use of the fourth-order CE can improve agreement with exact benchmarks (Fig. 1(b)) but can also lead to unphysical results associated with instabilities and negative spectral weight.  While carrying out the CE to infinite-orders yields exact results, it is clear that the manner in which convergence occurs is complicated.  Here, we focus on the time domain, explicitly illustrating how higher-order expansions \textit{always} systematically improve the accuracy of the short-time behavior.  To carry out this comparison, we employ the numerically exact ``Hierarchical Equations-of-Motion" (HEOM) method.\cite{Tanimura1989,Ishizaki2005,Tanimura2006,Ishizaki2009,Shi2009,Liu2014,Chen2015,Dunn2019}
This approach provides rapid convergence to the exact result for models such as the spin-boson model. For the one-dimensional Holstein model, exact convergence for finite times is attainable for weak-to-moderate coupling strengths in moderately-sized chains.\cite{Dunn2019}
Since this method may be unfamiliar to some readers, a brief description is provided in Appendix B.

We first work at weak coupling ($\omega_{0}=t_{0}=1, g=0.25$) and high temperatures ($T=1.0$) where we can easily converge the exact HEOM results for times sufficient to provide the full spectral function
with minimal artificial damping.
In Fig.~\ref{fig:heom_ce_small_g}(a) we show results for $N=6$ and $k=0$ which are consistent
with the behavior found in Fig. 1.  In particular, the fourth-order CE improves subtly on the second-order result, bringing the theory into
quantitative agreement with exact spectral function, with the exception of very small secondary peak structure visible at $\omega \sim -2$ and $\omega \sim 0$.  As expected from the discussion in Sec.IIA, this behavior is due to finite size recurrences which are expected to vanish as $N \rightarrow \infty$.  Fig.~\ref{fig:heom_ce_small_g}(b) illustrates the behavior underlying the spectral function in the frequency domain.  In particular, a large visible recurrence starting at $\omega_{0}t \sim 20$, which is missed by the second- and fourth-order CEs, can be observed.  

On the scale of Fig.~\ref{fig:heom_ce_small_g}(b), it is nearly impossible to parse what the fourth-order CE provides over the second-order CE to improve the distribution of spectral weight as seen in Fig.~\ref{fig:heom_ce_small_g}(a).  However, by focusing on the function $\Phi(k,t)$ directly, one clearly observes the systematic improvement provided by the fourth-order CE over the second-order CE.  In Table \ref{tab:table_1}, we consider two temperatures and two coupling 
strengths, along with the time, $t_\mathcal{M}$, after which $|\mathcal{G}_{\mathcal{M}}\left(t\right)-\mathcal{G}_{\infty}\left(t\right)|>5\times10^{-4}$,
where $\mathcal{M}$ is the order of the CE, and $\mathcal{G}_\infty$ is given by the exact HEOM result.
This improvement in short-time behavior is manifest in Fig. \ref{fig:heom_ce_small_g}(c), which illustrates the improved description of $\Phi(0,\omega)$ for parameters such that the long-time limit of the fourth-order CE is divergent.  In particular, in all cases, $t_{4}>t_{2}$ indicating that the fourth-order CE improves upon second-order CE.
Clearly, the long-time pathological behavior of the fourth-order CE does not corrupt the increase in accuracy of the short-time behavior of the cumulant generating function.
\begin{table}[tb]
    \centering
    \begin{ruledtabular}
    \begin{tabular}{l l c c}
        $g$    & $T/\omega_0$    & $t_2  \omega_0$   &  $t_4  \omega_0$ \\
        \hline
        0.25 & 0.0 & 2.90   & 8.70   \\
        0.25 & 1.0 & 1.35  &  5.55 \\
        1.0 & 0.0   &  0.50 &  0.85 \\
        1.0 & 1.0   &  0.35 & 0.65 
    \end{tabular}
    \end{ruledtabular}
    \caption{$t_{\mathcal{M}}$ at two EPI strengths and temperatures. 
    In all cases, raising either the temperature or the coupling decreases $t_{\mathcal{M}}$ and
    $t_2 < t_4$, indicating that the fourth-order CE improves upon the second-order CE at short times.
    Model parameters: 
    $N = 6; \omega_{0}=t_0=1$.
    HEOM calculations were performed using modified versions of 
  \textit{PHI}\cite{Strumpfer2012} and\textit{ pyrho.}\cite{Berkelbach} }
    \label{tab:table_1}
\end{table}

The above discussion suggests that convergence of the CE occurs in the time domain such that the short-time behavior can be systematically converged for longer and longer times, 
while concomitantly longer-time anomalies in $\Phi(k,t)$ must resum into functions which behave in a non-singular manner.  It is difficult to guess the form taken by such functions from just the first two terms in the expansion.  In this sense the fourth-order CE does not appear to be generically useful.  In the companion paper, we will present a self-consistent cumulant scheme that, while still suffering from some of the ill-effects introduced by the fourth-order CE, does provide access to non-perturbative behavior that appears to be completely out of reach of low-order CEs.\cite{RobinsonCumulantII}

\subsection{Thermodynamic Limit\label{sec:thermodynamic_limit}}

\begin{figure}[tb]
    \centering
    \includegraphics{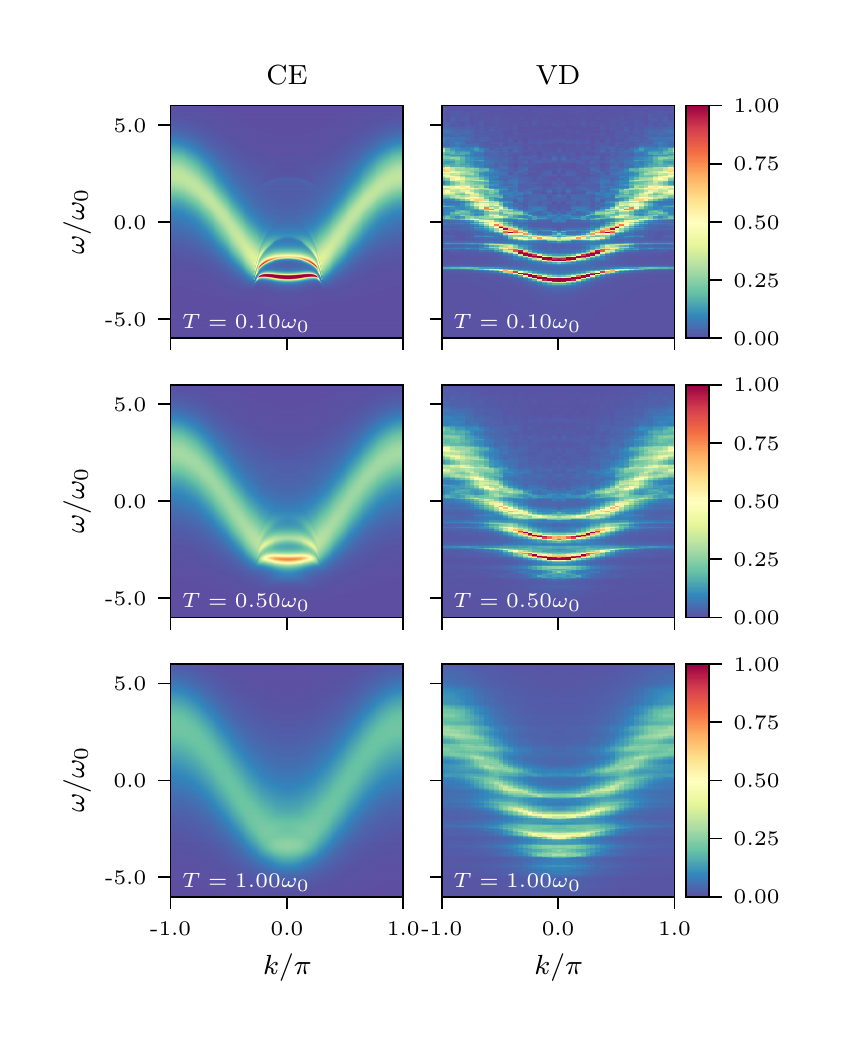}
    \caption{Heat maps of $A_2(k,\omega)$ for $N=600$ (left), and $A_{\infty/\text{VD}}(k,\omega)$ for $N=6$ (right, from Ref. \citenum{Bonca2019}) for several temperatures. 
    Note that for the CE there is a sharp transition at around $\pi/3$ from a clear quasiparticle peak to a incoherent spectrum.
    Additionally, note that compared to the VD results that the band curvature is qualitatively incorrect for the vibronic peaks at small $k$.
    Model parameters: $\omega_0=t_0=1, g^2 =2$, and $ \gamma=0.05$.
    }
    \label{fig:heat_maps}
\end{figure}

\begin{figure*}[tbh]
\begin{centering}
\includegraphics{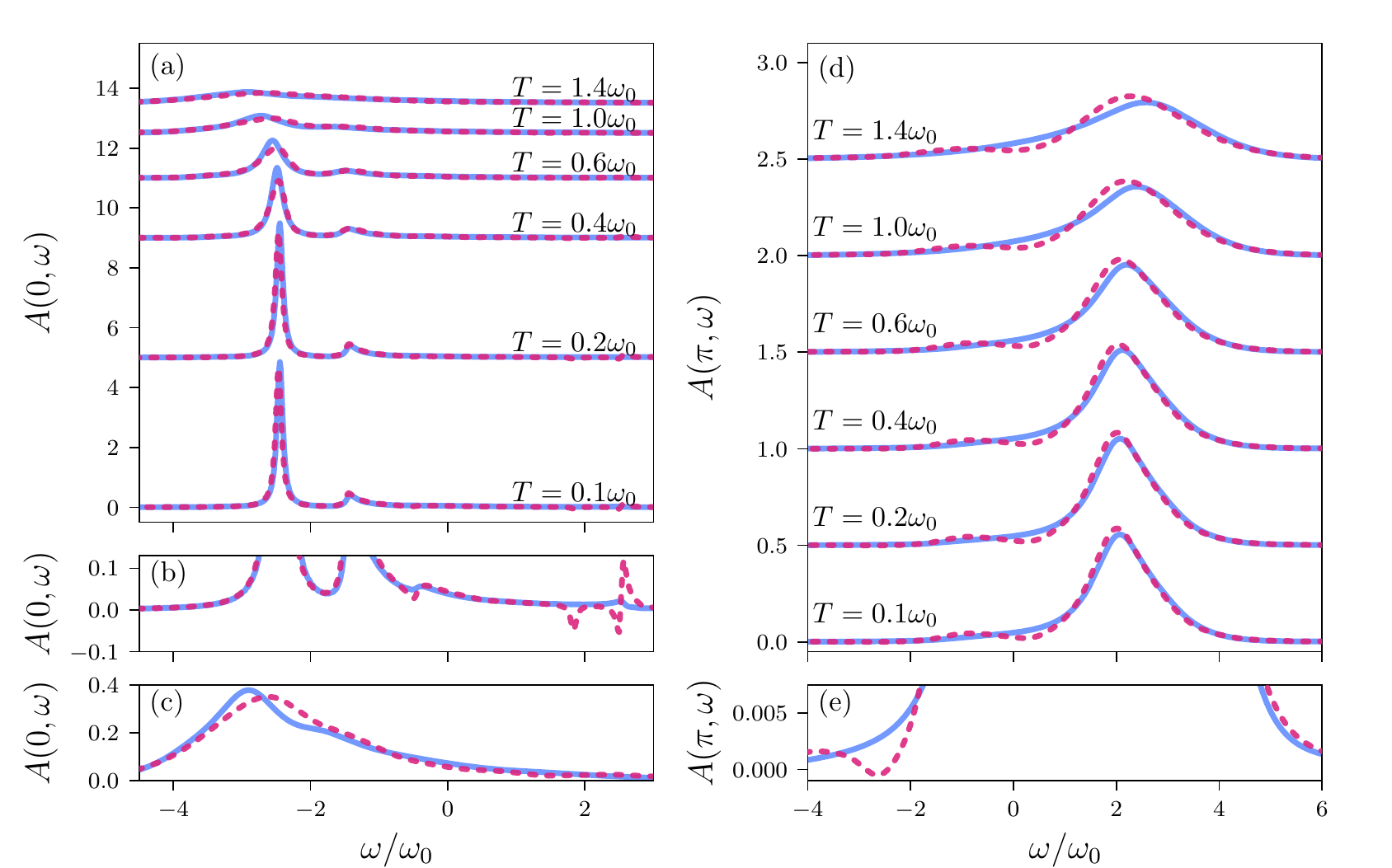}
\par\end{centering}
\caption{Spectral functions from the second-order CE (solid blue line) and the fourth-order CE (dashed pink line)
for $N=600$ Holstein model for a range of temperatures.
(a) Results for the band bottom $(k=0)$ for $T=[0.1,1.4].$ At low temperatures fourth-order CE slightly lowers the quasiparticle 
energy while also inserting regions of negative spectral weight at high frequencies.
(b) $k=0, T=0.1$ spectral functions enlarged to emphasize the negative spectral weight predicted by the fourth-order cumulant.
(c) $k=0, T=1.4$ spectral functions enlarged to emphasize the shift in peak location between the second- and fourth-order CE.
(d) Results for the band edge $(k=\pi)$ for $T=[0.1,1.4]$. The fourth-order CE prediction lowers the energy of the main peak predicted by second-order CE, and additionally adds a broad weak intensity peak at $-\omega_0$.
Both orders of the CE broaden similarly with increasing temperature.
(e) $k=\pi, T=0.1$ spectral functions enlarged to emphasize the added peak and the region of negative spectral weight predicted by the 
fourth-order CE compared to the second-order CE.
Model parameters: $\omega_{0}=t_0=g=1; \gamma=0.05$ ($k=0$ only).}
\label{fig:ce_thermodynamic}
\end{figure*}

Inspired by the possible suppression of physical fine structure and
the elimination of poorly behaved spurious oscillations in the infinite-system
limit, we now continue in the spirit of Dunn's continuum
calculation on the Fr{\"o}hlich model \cite{Dunn1975} to treat the finite-temperature infinite
Holstein model in the thermodynamic limit using the CE. 

We start by investigating the finite-$k$ behaviour of $A_2(k, \omega)$  and $A_4(k, \omega)$
in Fig. \ref{fig:heat_maps} where $A_n(k, \omega)$ 
denotes the $n^\text{th}$-order CE approximation.
Here, $N=600$ for the CE calculations while $N=6$ for the VD results. 
We expect that on the scale of these plots, finite size effects in the VD results are small as discussed in 
Ref. \citenum{Bonca2019}.
Note, however, the large changes that appear within $A_2(k,\omega)$ as $N$ is increased.
While $A_2(0, \omega)$ is accurate compared to the exact result,
even near $k=0$ the curvature of the bands that represent 
satellite peaks are described in a qualitatively incorrect manner. 
Further, while the finite size effects described in section \ref{sec:cumulant_finite_size} do vanish after an abrupt change of behaviour which occurs at $\epsilon_k = \omega_0$, they are
replaced with a single incoherent band centered on $\epsilon_k$.
Thus, the prominent satellite structure for $k>\omega_0$ in the exact results is completely absent in $A_2(k, \omega)$.

The fourth-order CE (not plotted in Fig. \ref{fig:heat_maps})
produces an $A(k, \omega)$ broadly similar to 
that of the second-order CE for
all $k$ which produce non-divergent fourth-order results in our approximation of the thermodynamic limit (N=600).
However, $A_4(k, \omega)$ does produce some notable differences from $A_2(k, \omega)$.
The fourth-order CE correctly predicts positive curvature of the satellite peaks at low temperatures near $k=0$, marking an improvement over the second-order cumulant result.
Along with this improvement, one obvious feature that emerges is small regions of negative spectral weight.
The appearance of negative spectral weight, discussed previously by 
Gunnarsson \textit{et al.}
\cite{Gunnarsson1994} and in the previous subsection, has also been noted in other studies.
\cite{Zuehlsdorff2019, Anda2016}
Though only present in high energy regions of the spectrum near $k=0$, 
the negative spectral weight in $A_4(k, \omega)$ appears at much lower energies, approximately at the location of VD quasiparticle energy, for $k=\pi$.
Finally, there are large regions of $k$ space for which divergent behaviour occurs.
It is possible that these regions become well-behaved for even larger system sizes,
however, we have no evidence that the fourth-order CE is
globally well behaved for $N=\infty$.

In Fig. \ref{fig:ce_thermodynamic}
we closely examine the temperature 
dependence of $A_2(k, \omega)$ and $A_4(k, \omega)$ at $k=0$ and $k=\pi$. 
At low temperatures,
the $k=0$ CE appears nearly converged, with $A_{2}\left(0, \omega\right)$
and $A_{4}\left(0, \omega\right)$ showing nearly identical behaviour around the quasiparticle peak.
On the other hand, at $k=\pi$ the low temperature fourth-order CE does not appear converged 
with respect to the second-order CE, and $A_2(\pi, \omega)$ deviates notably from $A_4(\pi, \omega)$ for $\omega > 0$. 
This distinction in performance at $k=0$ and $k\neq0$ is consistent with the overall comparison 
of $A_n(k, \omega)$ with results from VD.
In the companion paper, we will discuss how $A(k, \omega)$ for $k\neq0$ can be more accurately 
calculated from a self-consistent cumulant approach.\cite{RobinsonCumulantII}

For the $k=0$ case, the apparent convergence of the CE for some temperatures warrants more consideration, and thus we
devote the remainder of this subsection to a more detailed discussion of this case. 
The second- and fourth-order CE results match best for
high and low $T$. 
The low $T$ convergence of the main spectral features
is supported by the analysis in Sec. \ref{subsec:Convergence-of-Cumulant},
which shows how the CE in the time-domain $\mathcal{G}\left(0,t\right)$
breaks down only at high $T$, where the phonon occupation numbers
$N_{0}$ contribute to a growth in the magnitude of higher-order
cumulants. Meanwhile, for very high temperatures, Dunn's argument
that the long time behavior of
$\mathcal{G}_{2}\left(0,t\right)$ and $\mathcal{G}_{4}\left(0,t\right)$ may 
markedly differ with differences hidden by rapid damping such
that the resulting extremely broad spectral
functions may appear converged, is borne out.\cite{Dunn1975}
This extreme damping is seen for $T=1.4$ highlighted in Fig. \ref{fig:ce_thermodynamic}(c), where although the centroids of the main
second- and fourth-order CE peaks are displaced, the broadening makes
the high $T$ results appear converged. Such misleading convergence
behavior was also seen for the 6-site system in Sec. \ref{sec:time_convergence}.
At intermediate temperatures such as $T=0.6$, apparently neither
the low temperature real-time convergence illustrated in Sec. \ref{subsec:Convergence-of-Cumulant},
nor the high temperature damping behaviour discussed above is operative, such
that the CE results shown in Fig. \ref{fig:ce_thermodynamic} display a lack of
convergence for intermediate temperatures, where the second-order
CE polaron peak is considerably shifted from the fourth-order
CE polaron peak.
The shift of the polaron peak
to lower energies exhibited by $A_2(0,\omega)$ seems to be an artificial feature
that is corrected in $A_{4}\left(0, \omega\right)$ where the center of the polaron peak
appears fixed in location with respect to temperature.
The origin of the distinction between $A_2(0, \omega)$
and $A_4(0, \omega)$ is subtle.
An important approximation that
distinguishes the second-order and fourth-order CEs is the second-order assumption \cite{Dunn1975}
\begin{align}
\varepsilon\left(k+q_{1}+q_{2}\right) & \approx\varepsilon\left(k+q_{1}\right)+\varepsilon\left(k+q_{2}\right)-\varepsilon\left(k\right).
\end{align}
This approximation implies that the second-order CE overestimates the energy
of multiphonon processes, for example the consecutive emission of
two phonons with $q_{1}=q_{2}=\pi$.
This is likely the origin of the unphysical shift in $A_2(0, \omega)$. 

\section{Conclusion\label{sec:Conclusion}}
In this paper we have presented a comprehensive study of the behavior and properties of the CE method for one paradigmatic model of an electron interacting with phonons, namely the one-dimensional Holstein model.  The motivation for this choice is the fact that it is in this case where the most extensive numerically exact results are available for comparison.
Some of the conclusions we draw may be generic and connect to other polaron models, but future work is necessary before such a conclusion can be drawn.

Although of great interest for applications to realistic
systems, the CE in higher dimensional systems
is not directly explored in this work.
The formalism for the CE presented here is
dimensionality agnostic, so we expect that many of our
conclusions should hold in higher dimensions.

Within the confines of the second-order CE, we find that the spectral function is rather well described for up to intermediate coupling strengths at both low and high ($T>\omega_{0}$) temperatures for $k = 0$ but is quantitatively accurate for other wave vectors in the high temperature regime only.  Finite lattice effects are present in the numerically exact simulations which are not captured by low-order CE methods.  These features are small, and are not expected to be present in the infinite lattice limit.  The correct placement of satellite peaks for $k=0$ is revealed in the structure of the second-order CE in part to be the result of a spurious intensity crossing structure.  In the infinite size limit this structure is converted into a satellite region which exists only for wave vectors such that $(\epsilon_{k}+2t_0-\omega_{0})<0$ with {\em negative} band curvature for the higher order satellites.
The effect of other models and parameters on the size of this region is a topic of future study. Both the sharp change of behavior at $\epsilon_{k} + 2t_0 =\omega_{0}$ and the sign of the curvature contrast with the exact finite lattice results which are expected to semi-quantitatively describe the infinite lattice behavior. 

We have also explored the properties of the fourth-order CE.  At fourth order, we find that the short-time real-time evolution of the Green's function is always systematically improved, while the long-time behavior may become pathological depending on the parameters of the model and the wave vector in question.  When the fourth-order CE is well-behaved, improved spectral features are noted even at relatively low frequencies.  We have explored the origins of the ill-behaved fourth-order CE.  The general structure of the problematic terms take the algebraic form noted by Gunnarsson {\em et al.} to also give rise to negative spectral weight.~\cite{Gunnarsson1994}  In addition, classifying the divergent contributions for fixed lattice size N, we illustrate the subtle balance of terms that conspire to render the fourth-order CE either useful in correcting the second-order CE or pathological.  We note that in general the fourth-order CE does not generally appear capable of producing stable and sizable corrections to the second-order CE, even for intermediate electron-phonon coupling values.  In the companion paper, we formulate and study a self-consistent version of the CE which is capable of accurately capturing features beyond that of the low order CE.\cite{RobinsonCumulantII}

Lastly, we comment on the recent use of the CE for the study of transport behavior in real materials.  Specifically, Bernardi {et al.} have used the second-order CE, in conjunction with the ``bubble'' approximation to the current-current correlation function, to compute mobilities in both SrTi$O_{3}$ and in organic crystals. This approach has the advantage of capturing incoherent relaxation channels which are not described in the simplest semi-classical theories based on the Boltzmann equation. While we cannot comment on the accuracy of the CE for systems like SrTiO$_{3}$ for which the Fr\"{o}hlich model is most appropriate, nor can we comment on models with sizable Peierls coupling such as organic crystals, our results do suggest that the second-order CE should reasonably accurately model the full wave vector dependent one-particle spectral function for the situation $T \gtrsim \omega_{0}$, while likely becoming significantly less accurate for all but $k \sim 0$ at lower temperatures.  This of course does not imply that the independent bubble approximation is itself accurate.  Further work will be devoted to testing this approach in model systems where a controlled assessment of the various approximations is possible.

\section*{Acknowledgements}
The authors thank Prof. Janez Bon{\v c}a for providing the data from reference \citenum{Bonca2019}. 
P.J.R. acknowledges support from the National Science Foundation Graduate Research Fellowship under Grant No. DGE-2036197.
I.S.D. acknowledges support from the United States Department of Energy through the Computational Sciences Graduate Fellowship (DOE CSGF) under Grant No. DE-FG02-97ER25308. 
D.R.R. acknowledges support from NSF CHE-1954791.

\appendix
\section{Moments for the Holstein model\label{Appendix3}}

Plugging in the specific form of the Holstein interaction, performing the time integrals,
and removing several of the internal momentum sums via conservation
of momentum, we find the second moment is given by
\begin{align}
M_{2}&\left(k,t\right) =-\frac{g^{2}}{N}e^{-i\varepsilon_{k}t}\sum_{q}\biggr[\left(N_{0}+1\right)A_{q}^{-}+N_{0}A_{q}^{+}\biggr],\\
A_{q}^{\pm} & \equiv\frac{\pm it\left(\omega_{0}\pm\left(\varepsilon_{k}-\varepsilon_{q}\right)\right)-e^{\pm it\left(\omega_{0}\pm\left(\varepsilon_{k}-\varepsilon_{q}\right)\right)}+1}{\left(\omega_{0}\pm\left(\varepsilon_{k}-\varepsilon_{q}\right)\right)^{2}},
\end{align}
and the fourth moment is given by
\begin{align}
M_{4}(k,t) & =\frac{g^{4}}{N^{2}}e^{-i\varepsilon_{k}t}\sum_{q_{1},q_{2}}\nonumber \\
 & \biggr[\left(N_{0}+1\right)^{2}T_{1}\left(f_{1}^{+},f_{2}^{+};t\right)\nonumber \\
 & +N_{0}\left(N_{0}+1\right)T_{1}\left(f_{1}^{-},f_{2}^{+};t\right)\nonumber \\
 & +N_{0}\left(N_{0}+1\right)T_{1}\left(f_{1}^{+},f_{2}^{-};t\right)\nonumber \\
 & +N_{0}^{2}T_{1}\left(f_{1}^{-},f_{2}^{-};t\right)\nonumber \\
 & +\left(N_{0}+1\right)^{2}T_{2}\left(f_{1}^{+},f_{2}^{+},f\left(q_{1}+q_{2},0,2\omega_0\right);t\right)\nonumber \\
 & +N_{0}\left(N_{0}+1\right)T_{2}\left(f_{1}^{-},f_{2}^{+},f\left(q_{1}+q_{2},0,0\right);t\right)\nonumber \\
 & +N_{0}\left(N_{0}+1\right)T_{2}\left(f_{1}^{+},f_{2}^{-},f\left(q_{1}+q_{2},0,0\right);t\right)\nonumber \\
 & +N_{0}^{2}T_{2}\left(f_{1}^{-},f_{2}^{-},f\left(q_{1}+q_{2},0,-2\omega_0\right);t\right)\nonumber \\
 & +\left(N_{0}+1\right)^{2}T_{3}\left(f_{1}^{+},f\left(q_{1}+q_{2},0,2\omega_0\right);t\right)\nonumber \\
 & +N_{0}\left(N_{0}+1\right)T_{3}\left(f_{1}^{-},f\left(q_{1}+q_{2},0,0\right);t\right)\nonumber \\
 & +N_{0}\left(N_{0}+1\right)T_{3}\left(f_{1}^{+},f\left(q_{1}+q_{2},0,0\right);t\right)\nonumber \\
 & +N_{0}^{2}T_{3}\left(f_{1}^{-},f\left(q_{1}+q_{2},0,-2\omega_0\right);t\right)\biggr],\label{eq:-3.2}
\end{align}
where
\begin{align}
&T_{1}\left(a,b;t\right) =\frac{1}{b}\biggr[\frac{a\frac{t^{2}}{2}-t-h\left(a,t\right)}{a^{2}}\nonumber \\
 & +\frac{1}{b}\left(\frac{h\left(a,t\right)-h\left(b,t\right)}{a-b}
 -\frac{t+h\left(a,t\right)}{a}\right)\biggr],
\end{align}
\begin{align}
&T_{2}\left(a,b,c;t\right) =\frac{1}{b}\biggr[\frac{1}{c}\left(\frac{t+h\left(a,t\right)}{a}-\frac{h\left(a,t\right)-h\left(c,t\right)}{a-c}\right)\nonumber \\
 & -\frac{1}{c-b}\left(\frac{h\left(a,t\right)-h\left(b,t\right)}{a-b}-\frac{h\left(a,t\right)-h\left(c,t\right)}{a-c}\right)\biggr],\label{eq:-3.9}
\end{align}
\begin{align}
&T_{3}\left(a,b;t\right)  =\frac{1}{a}\biggr[\frac{1}{b}\left(\frac{t+h\left(a,t\right)}{a}-\frac{h\left(a,t\right)-h\left(b,t\right)}{a-b}\right)\nonumber \\
 & +\frac{1}{b-a}\left(\frac{e^{-at}t+h\left(a,t\right)}{a}+\frac{h\left(a,t\right)-h\left(b,t\right)}{a-b}\right)\biggr],\label{eq:-3.10}
\end{align}
\begin{align}
f(a,b,c) & =i(\varepsilon_{k+a}-\varepsilon_{k+b}+c),\\
f_{i}^{\pm} & =f\left(q_{i},0,\pm\omega_{0}\right),\\
h\left(x,t\right) & =\frac{e^{-xt}-1}{x}.
\end{align}

For both $M_{2}$ and $M_{4}$, singular terms within the momentum
sums are evaluated in a limiting sense using L'Hopital's rule.

\section{Hierarchical equations of motion\label{Appendix4}}

For an exact benchmark of $\mathcal{G}\left(k,t\right)$ we will
use the Hierarchical Equations of Motion (HEOM) approach.
First popularized for solving vibronic models with continuous bath
spectral densities, \cite{Tanimura1989,Ishizaki2005,Tanimura2006,Ishizaki2009,Shi2009}
HEOM has recently been adapted to solve discrete bath models such
as the Holstein and SSH models. \cite{Liu2014,Chen2015,Dunn2019}
While we have recently shown that the finite truncation of HEOM can lead
to long-time instability in such models, \cite{Dunn2019}
for the present application the converged short and intermediate time
behavior is sufficient to provide benchmarks for $\mathcal{G}\left(t\right)$
and $A\left(\omega\right)$.
Two recent versions of HEOM have provided practical routes to circumventing instabilities.\cite{Yan2020,jankovic2021spectral}

To compute $\mathcal{G}\left(k,t\right)$ with HEOM we rewrite
\begin{align}
\mathcal{G}(k,t) & =-i\Theta\left(t\right)\text{Tr}_{S}\big[a_{k}\text{Tr}_{B}\big[e^{-iHt}
\nonumber \\  & \times
\big(e^{-\beta H_{b}}\otimes a_{k}^{\dagger}\rho_{vac}\big)e^{iHt}\big]\big],
\end{align}
where 
\begin{align}
\rho_{vac} & =|0\rangle\langle0|=\left[\begin{array}{cccc}
1 & 0 & \dots & 0\\
0 & 0 & \dots & 0\\
\vdots & \vdots & \ddots & 0\\
0 & 0 & 0 & 0
\end{array}\right]
\end{align}
is the pure-state electronic density matrix representing the zero-electron
vacuum, written in a basis of zero-electron and one-electron states.
The $S$ and $B$ subscripts denote partial traces over the electron
and phonon subspaces, respectively. One-electron states are described
in the site basis. In this basis,
\begin{align}
a_{k}^{\dagger} & =\frac{1}{\sqrt{N}}\left[\begin{array}{cccc}
0 & 0 & \dots & 0\\
e^{-ik} & 0 & \dots & 0\\
\vdots & \vdots & \ddots & 0\\
e^{-ik\left(N-1\right)} & 0 & 0 & 0
\end{array}\right],\\
a_{k} & =\frac{1}{\sqrt{N}}\left[\begin{array}{cccc}
0 & e^{ik} & \dots & e^{ik\left(N-1\right)}\\
0 & 0 & \dots & 0\\
\vdots & \vdots & \ddots & 0\\
0 & 0 & 0 & 0
\end{array}\right].
\end{align}
Thus, to calculate $\mathcal{G}\left(k,t\right)$ we initialize a hierarchy
of auxiliary density matrices, each of dimension $\left(N+1\right)\times\left(N+1\right)$.
All matrices $\rho_{m_{1\pm},...,m_{N\pm}}(t=0)$ are set to zero
except for 
\begin{align}
\rho_{0,..,0}(t=0) & =a_{k}^{\dagger}\rho_{vac}~.
\end{align}
Then, we propagate in time using the discrete-bath HEOM \cite{Liu2014,Chen2015,Dunn2019}
\begin{align}
\frac{d}{dt}&\rho_{m_{1\pm},...,m_{N\pm}}(t)  =-i\mathcal{L}\rho_{m_{1\pm},...,m_{N\pm}}(t)\nonumber \\
 & -i\sum_{n=1}^{N}\omega_{0}\left(m_{n-}-m_{n+}\right)\rho_{m_{1\pm},...,m_{N\pm}}(t)\nonumber \\
 & +\sum_{n=1}^{N}\biggr[\Phi_{n}\biggr(\rho_{m_{1\pm},...,m_{n+}+1,...,m_{N\pm}}(t)\nonumber \\
 & +\rho_{m_{1\pm},...,m_{n-}+1,...,m_{N\pm}}(t)\biggr)\nonumber \\
 & +m_{n+}\Theta_{n+}\rho_{m_{1\pm},...,m_{n+}-1,...,m_{N\pm}}(t)\nonumber \\
 & +m_{n-}\Theta_{n-}\rho_{m_{1\pm},...,m_{n-}-1,...,m_{N\pm}}(t)\biggr],\label{eq:-3}
\end{align}
where
\begin{align}
\mathcal{L} & =[\hat{H}_{e},...],\\
\Phi_{n} & =[\hat{V}_{n},...]\\
\hat{V}_{n} & =a_{n}^{\dagger}a_{n},
\end{align}
and
\begin{align}
\Theta_{n\pm} & =-\frac{\left(g\omega_{0}\right)^{2}}{2}\left([\hat{V}_{n},...]\coth\left(\frac{\beta\omega_{0}}{2}\right)\mp\{\hat{V}_{n},...\}\right).
\end{align}
Finally, we compute the Green's function as
\begin{align}
\mathcal{G}(k,t) & =-i\Theta\left(t\right)Tr\left[a_{k}\rho_{0,..,0}(t)\right].
\end{align}
Converging with respect to the hierarchy depth $L,$ we obtain the
exact $\mathcal{G}\left(k,t\right)$ for the Holstein model.

\section{$K$-phonon approximation\label{Appendix5}}

For analyzing finite-size effects in an inexpensive, approximate way, we
will also compute $\mathcal{G}\left(k,t\right)$ via numerical diagonalization
of the Hamiltonian within a truncated basis. Toward this end we introduce
the momentum-space basis kets
\begin{equation}
|\nu_{0},\dots,\nu_{N}\rangle_{0},
\end{equation}
and
\begin{equation}
|k,\nu_{0},\dots,\nu_{N}\rangle_{1},
\end{equation}
which represent states with zero and one electron, respectively. The
electronic quantum number $k$ indicates the momentum of the electron.
The vibrational quantum numbers $\nu_{i}$ denote the number of vibrational
quanta in each normal mode, such that
\begin{align}
b_{q}^{\dagger}&|0,\dots,\nu_{q},\dots,0\rangle_{0} = \nonumber \\ & \sqrt{\nu_{q}+1}|0,\dots,\nu_{q}+1,\dots,0\rangle_{0},\\
b_{q}^{\dagger}& |k,0,\dots,\nu_{q},\dots,0\rangle_{1} = \nonumber \\ & \sqrt{\nu_{q}+1}|k,0,\dots,\nu_{q}+1,\dots,0\rangle_{1}.
\end{align}
We work within a truncated $K$ phonon basis such that
\begin{align}
\sum_{q=1}^{N}\nu_{q} & \leq K.
\end{align}
Using this basis to represent the Hamiltonian, we can then compute
the matrix exponential necessary to determine $\mathcal{G}\left(k,t\right)$
by numerically diagonalizing the Hamiltonian. We will refer to this
approach as the ``$K-$phonon approximation.''
In the text only $K=1$ results are shown.

\hspace{6em}

\bibliography{references_I}

\end{document}